\documentclass[10pt,aps,prb,twocolumn,preprintnumbers,amsmath,amssymb,floatfix,showpacs,citeautoscript,superscriptaddress]{revtex4-2}
\usepackage[latin1]{inputenc}
\usepackage[T1]{fontenc}
\usepackage[english]{babel}
\usepackage[pdftex]{graphicx}
\usepackage{amsmath}
\usepackage{times}
\usepackage[usenames,dvipsnames,svgnames,table]{xcolor}
\usepackage[colorlinks,plainpages=false,linkcolor=blue,urlcolor=blue,citecolor=blue,pdfpagemode=UseNone]{hyperref}
\usepackage{hyperref}

\renewcommand{\AA}{\text{\r{A}}}

\begin{document}

\title
{
\boldmath
Electronic reconstruction and interface engineering of emergent spin fluctuations in compressively strained La$_3$Ni$_2$O$_7$ on SrLaAlO$_4$(001)
}

\author{Benjamin Geisler}
\email{benjamin.geisler@ufl.edu}
\affiliation{Department of Physics, University of Florida, Gainesville, Florida 32611, USA}
\affiliation{Department of Materials Science and Engineering, University of Florida, Gainesville, Florida 32611, USA}
\author{James J. Hamlin}
\affiliation{Department of Physics, University of Florida, Gainesville, Florida 32611, USA}
\author{Gregory R. Stewart}
\affiliation{Department of Physics, University of Florida, Gainesville, Florida 32611, USA}
\author{Richard G. Hennig}
\affiliation{Department of Materials Science and Engineering, University of Florida, Gainesville, Florida 32611, USA}
\affiliation{Quantum Theory Project, University of Florida, Gainesville, Florida 32611, USA}
\author{P.J. Hirschfeld}
\affiliation{Department of Physics, University of Florida, Gainesville, Florida 32611, USA}

\date{\today}

\begin{abstract}
Motivated by the recent observation of ambient-pressure superconductivity with $T_c \sim 40$~K in La$_3$Ni$_2$O$_7$ on SrLaAlO$_4$(001)
\cite{Ko-LNO-ComprStrain-SC:25, Zhou-LNO-Strain:25},
we explore the structural and electronic properties as well as the spin-spin correlation function of this bilayer nickelate system by using density functional theory including a Coulomb repulsion term.
We find that the compressive strain exerted by this substrate leads to an unconventional occupation of the antibonding Ni~$3d_{z^2}$ states around the $\Gamma$ point,
distinct from the superconducting bulk compound under pressure.
While pure strain effects rather modestly enhance the dynamical spin susceptibility,
investigation of a reconstructed interface composition as observed in transmission electron microscopy
uncovers a strong amplification of the spin fluctuations  due to Fermi surface nesting of the antibonding Ni~$3d_{z^2}$ states near the interface.
These results provide insights into the emergence of superconductivity in strained La$_3$Ni$_2$O$_7$, suggest a possible key role of the interface, and highlight fundamental differences from the hydrostatic pressure scenario.
\end{abstract}

\maketitle

\section{Introduction}

The discovery of superconductivity with $T_c \sim 80$~K in pressurized La$_3$Ni$_2$O$_7$ \cite{Sun-327-Nickelate-SC:23, Hou-LNO327-ExpConfirm:23, Zhang-LNO327-ZeroResistance:23} has identified bilayer Ruddlesden-Popper nickelates as a fascinating new addition to the growing family of superconducting nickelates. This breakthrough has sparked significant theoretical and experimental interest \cite{Luo-LNO327:23, Gu-LNO327:23, Yang-LNO327:23, Lechermann-LNO327:23, Sakakibara-LNO327:23, Shen-LNO327:23, Christiansson-LNO327:23, Shilenko-LNO327:23, Wu-LNO327:23, Cao-LNO327:23, Chen-LNO327:23, Lu-LNO327-InterlayerAFM:23, ZhangDagotto-LNO327:23, Liao-LNO327:23, Qu-LNO327:23, HuangZhou-LNO327:23, QinYang-LNO327:23, Liu-LNO327-OxVacDestructive:23, ZhangDagotto-RE-LNO327:23, Liu-LNO327-Optics:23, Geisler-LNO327-Structure:23, RhodesWahl-LNO327:23, Wang-LNO327-SC-OxDeficient:23, Yang-LNO327-ARPES:23, Lu-LNO327:23, Wang-LNO327-SC:23, Wang-LNO327-I4mmm:23, Chen-LNO327-SDW:23, ZhengWu-LNO327:23, Kakoi-LNO327:23, Geisler-LNO327-Optical:24, LNO-327-Dong-VO-Visualization:24, Chen-Mitchell-Stacking-LNO327:24, Wang-La2PrNi2O7:24, LNO-327-SDW-LaBollita:24, Huo-LNO-Strain:25}.
Unlike infinite-layer nickelates, where superconductivity is limited so far to thin films \cite{Li-Supercond-Inf-NNO-STO:19, Li-NoSCinBulkDopedNNO:19, Botana-Inf-Nickelates:19, Li-Supercond-Dome-Inf-NNO-STO:20, Zeng-Inf-NNO:20, Osada-PrNiO2-SC:20, Wang-NoSCinBulkDopedNNO:20, Osada-LaNiO2-SC:21, GoodgeGeisler-NNO-IF:22}, La$_3$Ni$_2$O$_7$ exhibits superconductivity in bulk crystals.
The pairing mechanism has been linked to a pressure-driven Fermi surface reconstruction \cite{Sun-327-Nickelate-SC:23, Luo-LNO327:23, Gu-LNO327:23, Yang-LNO327:23, ZhangDagotto-LNO327:23} and an associated orthorhombic-to-tetragonal structural transition \cite{Geisler-LNO327-Structure:23, Wang-LNO327-I4mmm:23}, which suppresses octahedral rotations \cite{Sun-327-Nickelate-SC:23, Geisler-LNO327-Structure:23, ZhangDagotto-LNO327:23, Wang-LNO327-I4mmm:23}.
Depending on the balance of the involved Ni~$3d$ orbitals, it may correspond to $d$-wave or $s^{\pm}$ symmetry~\cite{Lechermann-LNO327:23, ZhangDagotto-LNO327:23, Liu-LNO327-OxVacDestructive:23, Xia-LNO327:25}.

Recently, epitaxial strain has been proposed by density functional theory as a promising route to induce ambient-pressure superconductivity in La$_3$Ni$_2$O$_7$~\cite{Geisler-LNO-Strained:24, Zhao-LNO-Strain:24, BhattaJia-LNO-Strain:25}.
Enhanced spin fluctuations have been predicted particularly for tensile strain in La$_3$Ni$_2$O$_7$/SrTiO$_3$(001)~\cite{Geisler-LNO-Strained:24}.
Intriguingly, independent experiments have since reported the observation of ambient-pressure superconductivity with a transition onset of up to $T_c \sim 40$~K in capped bilayer nickelate thin films of 4 to 6 bilayers grown on SrLaAlO$_4$(001) (SLAO) \cite{Ko-LNO-ComprStrain-SC:25, Liu-LNO-CompressiveStrain-SC:25, Zhou-LNO-Strain:25}, after previous attempts did not detect a superconducting transition for compressive or tensile strain~\cite{Cui-LNO-Strain:24}.
This necessitates %
the identification of potential differences to the physics of bilayer nickelates under high pressure
and how they relate to the pairing mechanism.

Here we explore the structural and electronic properties of La$_3$Ni$_2$O$_7$/SrLaAlO$_4$(001)
by using density functional theory including a Coulomb repulsion term,
with an explicit treatment of the interface.
In addition to the ideal interface stacking, we consider a reconstructed interface composition that has been identified in transmission electron microscopy (TEM)~\cite{Ko-LNO-ComprStrain-SC:25}.
Structural analysis reveals close agreement with recent x-ray diffraction (XRD) and TEM data
for La$_3$Ni$_2$O$_7$ on different prominent substrates~\cite{Bhatt-LNO-Strain:25}
and emphasizes the importance of the interfacial reconstruction.
We observe finite octahedral tilts throughout the bilayer nickelate,
which surprisingly are quenched at the reconstructed interface due to geometric constraints.
We find that compressive strain in the presence of a substrate induces an electronic reconstruction in La$_3$Ni$_2$O$_7$, characterized by an unconventional partial occupation of the antibonding Ni~$3d_{z^2}$ states.
Interestingly, while  compressive strain alone leads to a rather modest enhancement of the dynamical spin susceptibility,
the reconstructed interface composition substantially amplifies the spin fluctuations.
A detailed disentanglement of the scattering channels reveals that this phenomenon is driven by strong Fermi surface nesting involving the emergent interfacial Ni~$3d_{z^2}$ states.

These results suggest that the enhanced pairing in compressively strained La$_3$Ni$_2$O$_7$
arises from a different set of electronic excitations than the flat-band driven superconductivity in the bulk compound under hydrostatic pressure.
The pronounced sensitivity to the interface geometry highlights the potential of interface engineering in shaping the electronic properties and spin fluctuations in bilayer nickelates, opening new avenues for manipulating their superconducting phase.

\begin{figure*}
\begin{center}
\includegraphics[width=\linewidth]{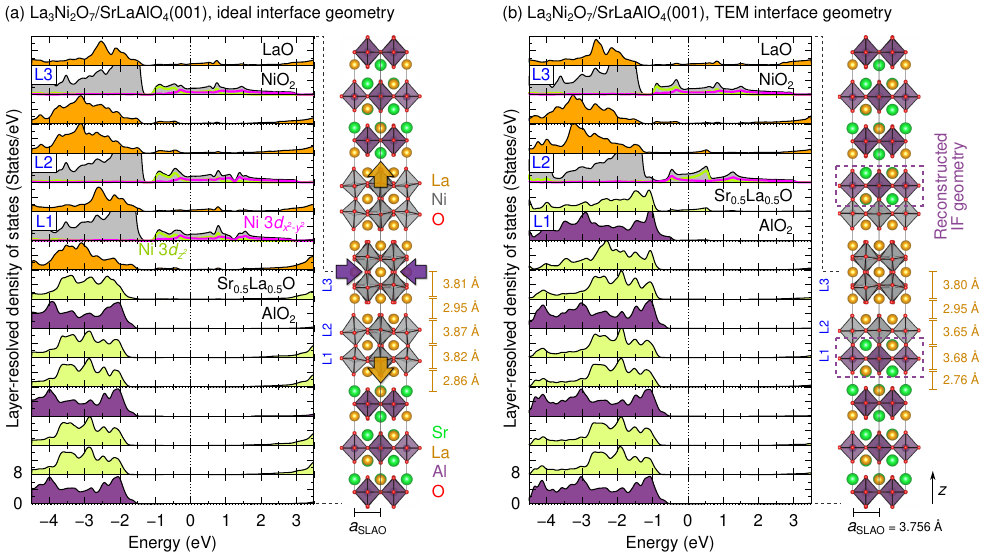}
\caption{\label{fig:OptGeo}
\textbf{\boldmath Optimized geometry and layer-resolved density of states of compressively strained La$_3$Ni$_2$O$_7$/SrLaAlO$_4$(001).}
(a)~Ideal interface stacking, representing a natural continuation of the two constituent oxides.
(b)~Interface composition as observed in recent TEM experiments~\cite{Ko-LNO-ComprStrain-SC:25},
consisting of an Al$_\text{Ni}$ substitution in layer~L1 and a mixed Sr$_{0.5}$La$_{0.5}$ configuration at the adjacent $A$ sites (purple dashed rectangles).
The orange numbers denote the $A$-site distances in $z$~direction.
Both systems show no involvement of the substrate in the Fermi surface,
although the interface reconstruction significancy reduces the energy difference between the Fermi level and the valence band maximum of the substrate.
}
\end{center}
\end{figure*}

\section{Results}

\subsection{\boldmath Interface geometry and electronic reconstruction in La$_3$Ni$_2$O$_7$/SrLaAlO$_4$(001)}

Figure~\ref{fig:OptGeo} shows the optimized geometry and layer-resolved density of states of 
epitaxial La$_3$Ni$_2$O$_7$ on SLAO(001).
We descibe the system by using symmetric $\sqrt{2}a \times \sqrt{2}a \times c$ supercells
that contain two equivalent interfaces (see Methods).
The substrate features a lattice constant of $a_\text{SLAO} = 3.756~\AA$~\cite{Ko-LNO-ComprStrain-SC:25, Bhatt-LNO-Strain:25}
and exerts $\varepsilon = a_\text{SLAO}/a_\text{LNO} - 1 \approx -1.9\%$ compressive strain on the bilayer nickelate,
which has a pseudocubic lattice parameter of $a_\text{LNO} = 3.83~\AA$.
We compare two representative interface geometries:
the ideal interface, which represents a natural continuation of the two constituent structures [Fig.~\ref{fig:OptGeo}(a)],
and a reconstructed geometry as suggested by recent TEM imaging~\cite{Ko-LNO-ComprStrain-SC:25}.
The latter consists in substituting Al for Ni ions in layer~L1, %
while the adjacent $A$ sites adopt a mixed Sr$_{0.5}$La$_{0.5}$ configuration instead of pure La ions,
which we treat explicitly here [Fig.~\ref{fig:OptGeo}(b)].

For the ideal interface geometry [Fig.~\ref{fig:OptGeo}(a)], the vertical $A$-site distances in the nickelate are enhanced to $3.82$-$3.87~\AA$ due to compressive strain, compared to the bulk value of $3.70~\AA$, and exhibit only minor variations.
In contrast, for the reconstructed interface [Fig.~\ref{fig:OptGeo}(b)], the $A$-site distance %
across layer~L2 is notably reduced to $3.65~\AA$ due to the proximity of the AlO$_2$ layer~L1.
We observe an expansion of the bilayer separation to $2.95~\AA$ in both systems, relative to the bulk value of $2.90~\AA$.

We identify finite octahedral tilts throughout the nickelate region in both optimized geometries (Fig.~\ref{fig:OptGeo}).
This contrasts with their assumed suppression in superconducting bulk La$_3$Ni$_2$O$_7$ under high pressure~\cite{Sun-327-Nickelate-SC:23, Geisler-LNO327-Structure:23, ZhangDagotto-LNO327:23, Wang-LNO327-I4mmm:23}.
Interestingly, a notable exception occurs in layer~L2 near the reconstructed interface, where the NiO$_6$ octahedral tilts are quenched.
This phenomenon is likely driven by the direct connectivity to the rigid, untilted AlO$_6$ octahedra in layer~L1,
which impose a geometric constraint on the adjacent nickelate layer.

\begin{figure*}
\begin{center}
\includegraphics[width=\linewidth]{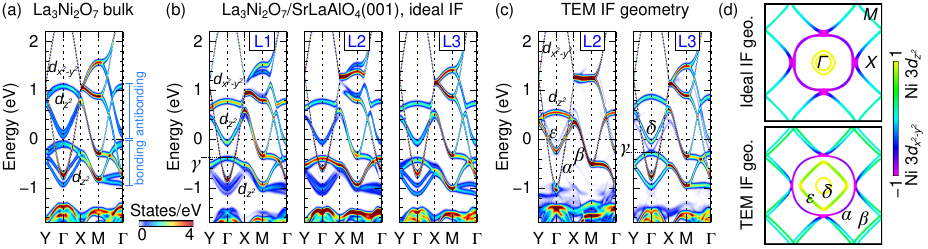}
\caption{\label{fig:Bands}
\textbf{\boldmath Electronic structure of the strained bilayer nickelate systems.}
Momentum-resolved density of states $A_i(\epsilon, k)$ (see text), projected on the Ni~$3d$ orbitals
(b)~in layers L1-L3 of La$_3$Ni$_2$O$_7$/SLAO(001) with ideal interface and
(c)~in layers L2-L3 of La$_3$Ni$_2$O$_7$/SLAO(001) with reconstructed interface (cf.~Fig.~\ref{fig:OptGeo}).
Consistently obtained Ni~$3d$ bands for bulk La$_3$Ni$_2$O$_7$ at ambient pressure are shown for comparison in panel~(a).
In the reconstructed system, Ni is substituted by Al in layer~L1;
hence, the adjacent layer~L2 presents a single Ni~$3d_{z^2}$ band instead of a bonding-antibonding pair.
(d)~Fermi surfaces in the $\sqrt{2}\times\sqrt{2}$ Brillouin zone, colored by the Ni orbital character $(3d_{z^2}-3d_{x^2-y^2})/(3d_{z^2}+3d_{x^2-y^2})$.
Intriguingly, compressive strain leads to the occupation of the \textit{antibonding} Ni~$3d_{z^2}$ states at the $\Gamma$~point ($\delta$ and $\varepsilon$ electron pockets),
in sharp contrast to the flat-band physics involving the $\gamma$ hole pocket in pressurized bulk La$_3$Ni$_2$O$_7$.
}
\end{center}
\end{figure*}

Along the [001] direction,
each bilayer of the 327 Ruddlesden-Popper nickelate consists of three (LaO)$^{1+}$ and two (NiO$_2$)$^{1.5-}$ layers,
yielding a nominal Ni~$3d^{7.5}$ valence.
The SLAO substrate, on the other hand,
is a band insulator with a formal Al$^{3+}$ valence state
and adopts a 214 Ruddlesden-Popper structure, %
consisting of single (AlO$_2$)$^{1-}$ layers
confined in two (Sr$_{0.5}$La$_{0.5}$O)$^{0.5+}$ layers.
At the ideal interface, the nickelate-substrate separation amounts to $2.86~\AA$, %
while it is contracted to $2.76~\AA$ at the reconstructed interface (Fig.~\ref{fig:OptGeo}).
This reduction can be directly attributed to the lowered formal polarity of the $A$O layers involved.
For reference, La$_3$Ni$_2$O$_7$ on non-polar SrTiO$_3$(001) presents a similar interfacial $A$-site distance of $2.78~\AA$~\cite{Geisler-LNO-Strained:24}.

The layer-resolved densities of states in Fig.~\ref{fig:OptGeo} show the relative band alignment of the constituent oxides.
We find in both systems that the Fermi energy is clearly located in the substrate band gap.
Therefore, the Fermi surfaces shown in Fig.~\ref{fig:Bands}(d) are exclusively formed by the Ni~$e_g$ states.
The interface reconstruction reduces the energy difference between the Fermi level and the valence band maximum of the substrate from $\sim 1.5$ to $\sim 0.6$~eV.
In contrast to La$_3$Ni$_2$O$_7$/LaAlO$_3$(001),
where the polar discontinuity at the interface leads to an electron doping of the nickelate superconductor~\cite{Geisler-LNO-Strained:24},
we do not observe interfacial charge transfer for the present systems.

Figure~\ref{fig:Bands} shows the projected momentum-resolved density of states,
$A_i(\epsilon, k) = \sum_{n} \, \langle \phi_i \vert \psi_{n,k} \rangle \, \delta(\epsilon_{n,k}-\epsilon)$,
where $\phi_i$ denotes Ni~$3d$ manifolds in different layers~$i$. %
In general, compressive strain drives a significant charge transfer from the Ni~$3d_{x^2-y^2}$ reservoir to the Ni~$3d_{z^2}$ states,
since the concomitant vertical expansion of the nickelate region tends to lower the energy of the Ni~$3d_{z^2}$ orbital relative to Ni~$3d_{x^2-y^2}$.
Intriguingly, this allows to control the Ni~$e_g$ orbital polarization
$(3d_{z^2}-3d_{x^2-y^2})/(3d_{z^2}+3d_{x^2-y^2})$
over a considerably larger interval than hydrostatic pressure~\cite{Geisler-LNO-Strained:24}.
Here, we find that this mechanism leads to an unconventional occupation of the \textit{antibonding} Ni~$3d_{z^2}$ states around the $\Gamma$ point [Fig~\ref{fig:Bands}(b)],
which are empty in the bulk [Fig~\ref{fig:Bands}(a)], even at finite pressure.
Their occupation is fairly uniform throughout the different layers, signaling the absence of electrostatic doping.
Simultaneously, the \textit{bonding} Ni~$3d_{z^2}$ states are entirely filled
and their flat-band-like maximum is universally lowered to $\sim -0.37$~eV.

Importantly, this picture applies to the system with reconstructed interface as well [Fig~\ref{fig:Bands}(c)].
We observe the $\gamma$ flat band now at $\sim -0.26$~eV in layer~L3.
Due to the Al$_\text{Ni}$ substitution in layer~L1,
the adjacent layer~L2 presents a single Ni~$3d_{z^2}$ band ($\varepsilon$) instead of a bonding-antibonding pair. %
It is more strongly filled than the other Ni~$3d_{z^2}$ bands in either the ideal or the reconstructed system
and provides a substantial contribution to the total density of states at the Fermi energy [Fig.~\ref{fig:OptGeo}(b)],
which plays a central role in the enhancement of the spin susceptibility (see below).
Analysis of the layer polarities leaves the NiO$_2$ layer~L2 in a formal ${1-}$ state,
leading to an altered local Ni$^{3+}$ valence similar to bulk LaNiO$_3$.
Interestingly, this interfacial NiO$_2$ layer is embedded in a highly anisotropic chemical environment, which induces a considerable Ni~$e_g$ orbital polarization.
This effect is even more pronounced than in LaNiO$_3$/LaAlO$_3$(001) superlattices, which have been widely explored in the context of
Fermi surface engineering and nickelate superconductivity~\cite{ChaloupkaKhaliullin:08, Hansmann:09}.

The Fermi surface of La$_3$Ni$_2$O$_7$ is known to consist of two characteristic sheets:
an $\alpha$ sheet of predominantly Ni~$3d_{x^2-y^2}$ character
and a $\beta$ sheet with some admixture of Ni~$3d_{z^2}$ \cite{Luo-LNO327:23, Gu-LNO327:23, Yang-LNO327:23, Yang-LNO327-ARPES:23}.
Under hydrostatic pressure~\cite{Sun-327-Nickelate-SC:23, Luo-LNO327:23, Gu-LNO327:23, Yang-LNO327:23, Lechermann-LNO327:23, Liu-LNO327-OxVacDestructive:23, ZhangDagotto-LNO327:23} or tensile strain as imposed by SrTiO$_3$(001)~\cite{Geisler-LNO-Strained:24},
the system undergoes a topological transition, i.e., the emergence of the $\gamma$ hole pocket due to a metallization of the bonding Ni~$3d_{z^2}$ states.
In stark contrast, our simulations for compressively strained La$_3$Ni$_2$O$_7$/SLAO(001) reveal a distinct Fermi surface composition,
reflecting the differences in electronic reconstruction [Fig.~\ref{fig:Bands}(d)].
The $\beta$ sheet exhibits an enhanced Ni~$3d_{z^2}$ contribution compared to the bulk, while the $\alpha$ sheet appears more rectangular and features a significantly promoted Ni~$3d_{x^2-y^2}$ character.
Most importantly, electron pockets emerge near the $\Gamma$ point, originating from the partial occupation of the antibonding Ni~$3d_{z^2}$ states.
Specifically, for the reconstructed interface composition, we find a slight splitting of the $\beta$ sheet,
a larger electron pocket $\varepsilon$ around the $\Gamma$ point predominantly associated with Ni~$3d_{z^2}$ states localized in layer~L2, and a smaller pocket $\delta$ that stems from layer~L3 [Fig.~\ref{fig:Bands}(c,d)].

The pronounced effect of compressive strain and interface reconstructions on the electronic structure suggests their critical role in modulating the superconducting properties.
We will corroborate this aspect by calculating the dynamical spin susceptibility below.

\subsection{Structural properties of bilayer nickelates on different substrates}

\begin{figure}
\begin{center}
\includegraphics[width=\linewidth]{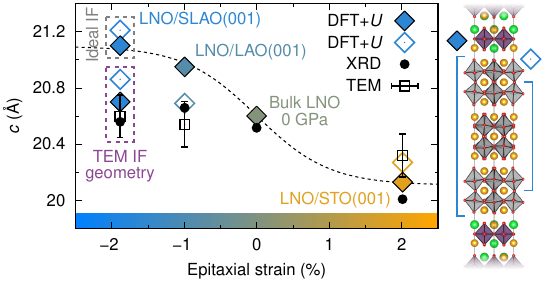}
\caption{\label{fig:Structure}
\textbf{\boldmath Structural properties of bilayer nickelates on different substrates.} Evolution of the effective $c_\text{LNO}$ of La$_3$Ni$_2$O$_7$ (LNO) as a function of the epitaxial strain~$\varepsilon$.
For the DFT+$U$-optimized structures reported here and in previous work~\cite{Geisler-LNO-Strained:24},
we show $c_\text{LNO}$ measured in two different ways as described in the text (filled and empty diamonds; see structure on the right).
The results are compared to recent XRD and TEM measurements for strained La$_3$Ni$_2$O$_7$ thin films~\cite{Bhatt-LNO-Strain:25} and XRD for bulk LNO at ambient pressure~\cite{LNO-327-Diffraction-Zhang:94}.
}
\end{center}
\end{figure}

The availability of high-quality x-ray diffraction and TEM data on strained epitaxial thin films~\cite{Bhatt-LNO-Strain:25},
specifically for the evolution of the vertical lattice parameter of the bilayer nickelate,
offers a unique opportunity to systematically compare experimental observations with the optimized structures reported here and in previous work~\cite{Geisler-LNO-Strained:24}.
Extracting an effective nickelate lattice parameter $c_\text{LNO}$ from the optimized structures can be performed in different ways,
which yield slightly different values.
In Fig.~\ref{fig:Structure}, we present $c_\text{LNO}$ obtained by using two distinct methods:
(i)~The \textit{outer} distance (filled diamonds) is defined as the distance between the two interfaces.
This corresponds to $3/2 \cdot c_\text{LNO}$ of an orthorhombic La$_3$Ni$_2$O$_7$ reference cell.
(ii)~The \textit{inner} distance (empty diamonds) corresponds to $2/2 \cdot c_\text{LNO}$ and captures more specifically the internal structural response of the nickelate layer.

Interestingly, we find that the outer distance follows approximately a $\tanh(\varepsilon)$ function (dashed black line),
a behavior observed also in LaNiO$_3$/LaAlO$_3$(001) superlattices~\cite{GeislerPentcheva-LNOLAO-Resonances:19}.
For La$_3$Ni$_2$O$_7$/SLAO(001), it yields $c_\text{LNO} = 21.10~\AA$ (ideal interface) and $20.70~\AA$ (reconstructed interface).
Both values clearly exceed the bulk reference of $20.6~\AA$~\cite{Geisler-LNO327-Structure:23},
indicating a significant strain-induced vertical expansion of the nickelate.
The inner distance, which generally renders slightly larger values, yields $c_\text{LNO} = 21.21~\AA$ (ideal interface) and $20.86~\AA$ (reconstructed interface).
We observe that the reconstructed interface geometry reduces $c_\text{LNO}$ considerably, bringing the theoretical predictions for La$_3$Ni$_2$O$_7$/SLAO(001) into close agreement with the experimental TEM and XRD results (Fig.~\ref{fig:Structure}).
For La$_3$Ni$_2$O$_7$/SrTiO$_3$(001) (LNO/STO), we obtain very good agreement between theoretical and experimental values as well.
For La$_3$Ni$_2$O$_7$/LaAlO$_3$(001) (LNO/LAO), the inner distance aligns closely with the experimental data, whereas the outer distance is notably higher,
reflecting the complex interplay of interfacial charge transfer and structural distortions in this system.

Overall, our DFT$+U$ predictions agree within a $\pm 0.8\%$ margin with the experimental lattice constants
over a wide range of strain values $\varepsilon$ from $-2\%$ to $2\%$,
despite systematic uncertainties in the theoretical and experimental data.
This comparison underscores the impact of interface effects on the structural parameters.
The values compiled in Fig.~\ref{fig:Structure} will be a helpful reference for future experimental studies.

\begin{figure*}
\begin{center}
\includegraphics[width=\textwidth]{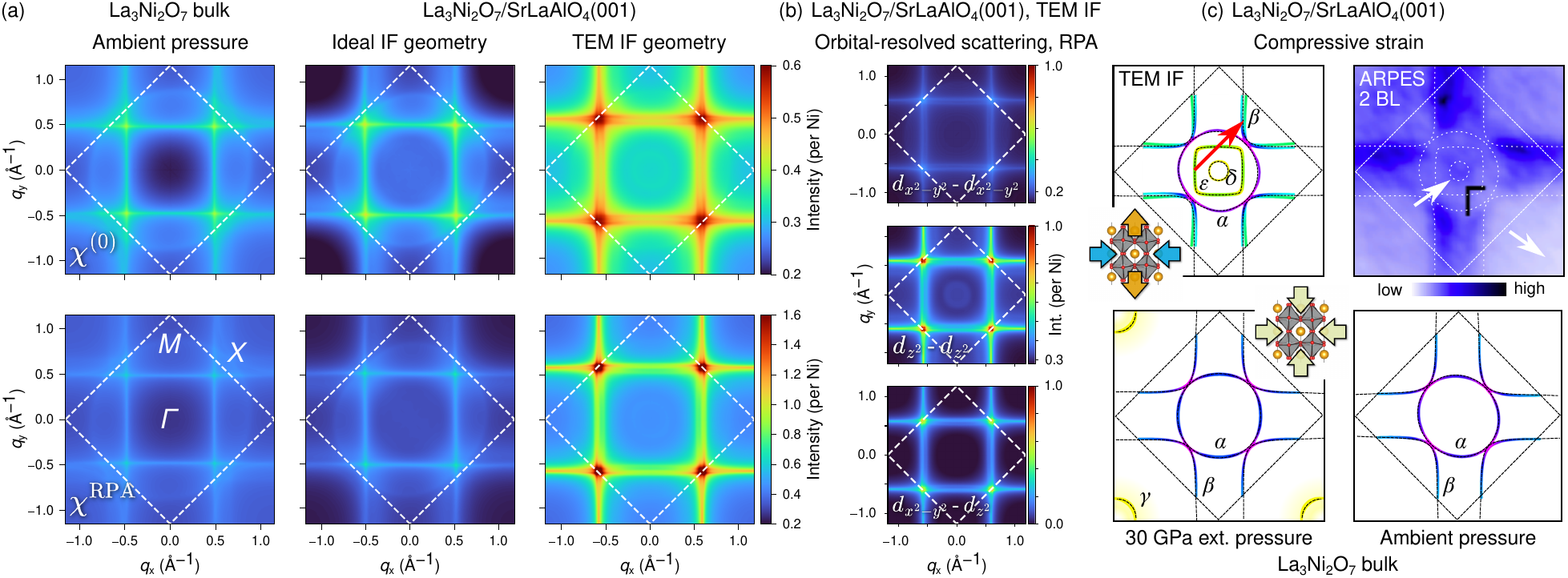}
\caption{\label{fig:Chi}
\textbf{\boldmath Spin-spin correlation functions and fundamental difference between strain and pressure in bilayer nickelates.}
(a)~Comparison of $\chi^{(0)}(\mathbf{q})$ and $\chi^\text{RPA}(\mathbf{q})$ for different systems, unfolded from the $\sqrt{2}\times\sqrt{2}$ (dashed lines) to the full Brillouin zone.
An enhanced magnitude and sharpness of the peaks along the $(\pi,\pi)$ direction can be observed in the strained systems
with respect to bulk La$_3$Ni$_2$O$_7$ at ambient pressure,
particularly in the case of the experimentally observed interface reconstruction.
(b)~The orbital-resolved contributions to $\chi^\text{RPA}(\mathbf{q})$
for La$_3$Ni$_2$O$_7$/SLAO(001) with reconstructed interface
are largest for the intraorbital $3d_{z^2}$ and interorbital $3d_{x^2-y^2}$-$3d_{z^2}$ channels.
(c)~Fermi surfaces, interpreted in the unfolded Brillouin zone.
The red nesting vector couples the $\beta$ and $\varepsilon$ sheets at the interface, both showing strong $3d_{z^2}$ contribution (Fig.~\ref{fig:Bands}),
and drives the enhancement of the $(\pi,\pi)$ peaks.
The panels illustrate the fundamentally different impact of compressive strain versus hydrostatic pressure, i.e.,
the formation of electron pockets around the $\Gamma$ point ($\delta$, $\varepsilon$)
versus flat-band hole pockets forming in the Brillouin zone corner ($\gamma$).
Recent ARPES measurements of strained samples on SLAO(001) showing an ambient-pressure superconducting transition (reproduced from Li \textit{et al.}~\cite{LNO327-SLAO-ARPES-Li:25}) confirm the suppression of the $\gamma$ corner pocket and indicate the appearance of a finite signal near the $\Gamma$ point (white arrows). The superimposed white lines correspond to the Fermi surface obtained here.
}
\end{center}
\end{figure*}

\subsection{\boldmath Spin susceptibility in La$_3$Ni$_2$O$_7$/SrLaAlO$_4$(001)}

The superconducting pairing is determined by the dynamical spin susceptibility $\chi^\text{RPA}(\mathbf{q})$,
which we compare for different systems in Fig.~\ref{fig:Chi}(a)
considering a Kanamori-type interaction vertex with $U=0.6$ and $J=0.2$~eV in the random phase approximation (RPA).
For completeness, we also show the bare Lindhard susceptibility $\chi^{(0)}(\mathbf{q})$.

The spin susceptibility of bulk La$_3$Ni$_2$O$_7$ at ambient pressure is dominated by intraorbital $3d_{x^2-y^2}$ scattering~\cite{Geisler-LNO-Strained:24}.
Direct comparison reveals that La$_3$Ni$_2$O$_7$/SLAO(001) exhibits a moderately enhanced magnitude and sharpness of the characteristic susceptibility peaks along the $(\pi,\pi)$ direction [Fig.~\ref{fig:Chi}(a)].
The difference can be traced back to the intraorbital $3d_{z^2}$ and interorbital $3d_{x^2-y^2}$-$3d_{z^2}$ channels (not shown).

Intriguingly, we find a very strong enhancement of the susceptibility for La$_3$Ni$_2$O$_7$/SLAO(001) if we consider the reconstructed interface geometry observed in TEM [Fig.~\ref{fig:Chi}(a)].
These changes are not only quantitative, but also qualitative,
as they are driven by considerably increased intraorbital $3d_{z^2}$ scattering
and further promoted by contributions from the interorbital $3d_{x^2-y^2}$-$3d_{z^2}$ channel [Fig.~\ref{fig:Chi}(b)].
Both clearly surpass the $3d_{x^2-y^2}$ channel.

This surprising enhancement of spin fluctuation weight
originates predominantly from the emergent Ni~$3d_{z^2}$ states localized near the interface (i.e., the $\varepsilon$ sheet),
which contribute strongly to the density of states at the Fermi level (Figs.~\ref{fig:OptGeo} and~\ref{fig:Bands})
and exhibit favorable nesting properties %
with the $\beta$ sheets, as indicated by the red vector $\sim (\pi,\pi)$ in Fig.~\ref{fig:Chi}(c).
Together, these effects result in the strong amplification of the spin fluctuations. %
Considering that recent experiments reporting ambient-pressure superconductivity in strained systems are performed on capped bilayer nickelate films of 4 to 6 bilayers thickness~\cite{Ko-LNO-ComprStrain-SC:25, Liu-LNO-CompressiveStrain-SC:25, Zhou-LNO-Strain:25},
these findings suggest that interfacial effects may play a key role in the emergence of superconductivity in La$_3$Ni$_2$O$_7$ on SLAO(001).

\section{Discussion}

We investigated the effect of compressive strain and interface reconstructions in La$_3$Ni$_2$O$_{7}$ on SrLaAlO$_4$(001)
by performing first-principles simulations including a Coulomb repulsion term
and explicitly treating the interface.
We found that compressive strain drives an electronic reconstruction in the bilayer nickelate,
characterized by an unconventional occupation of the antibonding Ni~$3d_{z^2}$ states
despite the absence of interfacial charge transfer.
This scenario contrasts sharply with the notion of metallized bonding Ni~$3d_{z^2}$ states,
considered to be critical for the superconducting state in pressurized bulk La$_3$Ni$_2$O$_7$.
The concomitant substantial modifications of the Fermi surface and amplified spin fluctuations,
especially for the reconstructed interface due to strong nesting of the cuprate-shaped $\beta$ sheets with the emergent Ni~$3d_{z^2}$ states,
point to an electron pairing that deviates fundamentally from the flat-band physics dominant under hydrostatic pressure,
as illustrated in Fig.~\ref{fig:Chi}(c).
We provide a tight-binding model (see Methods) that allows a detailed exploration of the gap structure and symmetry of the order parameter in future work.

Interestingly, recent angle-resolved photoemission spectroscopy (ARPES) experiments
published after initial submission of this paper
identified a suppressed contribution of the bonding Ni~$3d_{z^2}$ flat band (i.e., the $\gamma$ hole pocket) to the Fermi surface \cite{LNO327-SLAO-ARPES-Li:25, LNO327-SLAO-ARPES-WangStanford:25, LNO327-SLAO-ARPES-Shen:25}, with the respective states shifted to at least $\sim -70$~meV below the Fermi level \cite{LNO327-SLAO-ARPES-WangStanford:25}.
This behavior is in stark contrast to that of pressurized bulk samples and in qualitative agreement with our findings, as summarized in Fig.~\ref{fig:Chi}(c).
Moreover, it has been suggested that conduction is localized at the interface \cite{LNO327-SLAO-ARPES-Li:25}, implying a layer dependence of the electronic structure throughout the sample, as also found by our simulations treating realistic interface compositions.
Although the imaging resolution is limited so far, these findings raise a critical question:
If the bonding Ni~$3d_{z^2}$ states are not active at the Fermi level, how does the electronic structure of compressively strained La$_3$Ni$_2$O$_7$ differ from that at ambient pressure, and why does superconductivity emerge?
Suggesting a tentative answer, we show that electron pockets formed by antibonding Ni~$3d_{z^2}$ states around the $\Gamma$ point may play a relevant role and that interface-related phenomena---such as Sr and Al interdiffusion into the bilayer stacking, as experimentally observed \cite{Ko-LNO-ComprStrain-SC:25, LNO327-SLAO-ARPES-Li:25}---possibly enhance superconductivity.
Recent thin-film ARPES data provide indications that support this hypothesis \cite{LNO327-SLAO-ARPES-Li:25, LNO327-SLAO-ARPES-WangStanford:25}, as %
illustrated in Fig.~\ref{fig:Chi}(c), which calls for further investigation.

Comparison of the structural optimization results,
specifically the evolution of the vertical lattice parameter of the bilayer nickelate on different technologically relevant substrates,
with recent x-ray diffraction and transmission electron microscopy measurements on strained epitaxial thin films
revealed close agreement
and further corroborated the key role of the reconstructed interface composition.
We observed finite octahedral tilts throughout the bilayer nickelate.
Intriguingly, these tilts are suppressed at the reconstructed interface, accompanied by significantly enhanced spin fluctuations.
This finding provides an important contribution to the ongoing debate on whether superconductivity in bilayer nickelates requires a structural transition to the high-symmetry $I4/mmm$ phase or can arise from purely electronic effects~\cite{Geisler-LNO327-Structure:23, Geisler-LNO-Strained:24}.

The pronounced influence of epitaxial strain and interface effects underscores the potential of strain and interface engineering to tailor the electronic structure and optimize the superconducting properties in bilayer nickelates at ambient pressure,
circumventing the challenges associated with high-pressure synthesis and characterization and offering a pathway towards realizing their full technological potential.

\section{Methods}

\subsection{Density functional theory calculations}

We performed first-principles calculations in the framework of density functional theory~\cite{KoSh65} (DFT)
as implemented in the Quantum Espresso code~\cite{PWSCF}.
The generalized gradient approximation was used for the exchange and correlation functional as parametrized by Perdew, Burke, and Ernzerhof~\cite{PeBu96}
in conjunction with ultrasoft pseudopotentials~\cite{Vanderbilt:1990},
as successfully employed in previous work~\cite{GeislerPentcheva-LNOLAO:18, WrobelGeisler:18, GeislerPentcheva-LNOLAO-Resonances:19}.
Static correlation effects were considered within the DFT$+U$ formalism~\cite{QE-LDA-U:05},
employing $U=4$~eV on the Ni and Ti $3d$ states~\cite{GeislerPentcheva-InfNNO:20, GeislerPentcheva-NNOCCOSTO:21, Geisler-VO-LNOLAO:22, Sun-327-Nickelate-SC:23, ZhangDagotto-LNO327:23, Geisler-LNO-Strained:24}.
Octahedral rotations are fully accounted for by using
$\sqrt{2}a \times \sqrt{2}a \times c$ supercells with two transition metal sites per layer,
setting $a$ to the substrate lattice parameter of $a_\text{SLAO} = 3.756~\AA$~\cite{Ko-LNO-ComprStrain-SC:25, Bhatt-LNO-Strain:25}
and accurately optimizing $c$ as well as the ionic positions.
The symmetric supercells contain two equivalent interfaces, 3 bilayers of La$_3$Ni$_2$O$_7$, and 5 layers of substrate,
corresponding to 142 atoms (Fig.~\ref{fig:OptGeo}). %
This setup ensures that key interfacial effects are fully captured. Additional inner layers would primarily reflect bulk-like behavior and would not alter the essential physics.
Wave functions and density were expanded into plane waves up to cutoff energies of $35$ and $350$~Ry, respectively.
The Brillouin zone was sampled by using
a $12 \times 12 \times 1$ Monkhorst-Pack $\Vec{k}$-point grid~\cite{MoPa76}
and $5$~mRy Methfessel-Paxton smearing~\cite{MePa89}.
The ionic positions were accurately optimized, reducing ionic forces below $1$~mRy$/$a.u.
Subsequently, Fermi surfaces and densities of states were obtained by using \mbox{$64\times64\times4$} $\Vec{k}$-point grids.
Reference calculations for ambient-pressure bulk La$_3$Ni$_2$O$_7$ ($Cmcm$) were performed %
by using 48-atom supercells~\cite{Geisler-LNO327-Structure:23, Geisler-LNO327-Optical:24}
with \mbox{$12\times12\times4$} and \mbox{$64\times64\times8$} $\Vec{k}$-point grids.

\subsection{Tight-binding Hamiltonians in a basis of maximally localized Wannier functions}

We constructed tight-binding Hamiltonians describing the physics of the Ni~$e_g$ states,
which are the active states near the Fermi level,
in a basis of maximally localized Wannier functions
for all bilayer nickelate systems considered in this work.
The tight-binding models describe 12 (8) distinct Ni sites and thus 24 (16) Ni orbitals
for the system with ideal (reconstructed) interface,
which can be further reduced by exploiting symmetries.
The output files of \textit{wannier90}~\cite{Wannier90:20},
which can be directly imported into different post-processing programs,
are conveniently provided at \url{https://github.com/henniggroup}.

\subsection{Spin susceptibility calculations}

Using these tight-binding Hamiltonians,
we computed the bare generalized susceptibility $\chi^{(0)}_{abcd}(\mathbf{q})$
at \mbox{$T = 60$~K} ($\beta = 200$)
leveraging the Toolbox for Research on Interacting Quantum Systems-Two-Particle Response Function (TRIQS-TPRF) framework~\cite{TRIQS:15}
by Fourier transform of
\begin{equation}
\chi^{(0)}_{abcd}(\mathbf{r}) =
     - \int_0^\beta d\tau \,
     G^{(0)}_{da}(\tau, \mathbf{r}) G^{(0)}_{bc}(-\tau, -\mathbf{r}) 
\end{equation}
from non-interacting single-particle Green's functions defined on a mesh of Matsubara frequencies
\begin{equation}
G^{(0)}_{ab}(\mathbf{k}, i\omega_n) =
      \left[ i\omega_n \cdot \mathbf{1} - \epsilon(\mathbf{k}) \right]^{-1} \ ,  
\end{equation}
where $\epsilon_{ab}(\mathbf{k})$ is the matrix-valued dispersion relation.
The indices label the different orbitals ($3d_{z^2}$ and $3d_{x^2-y^2}$) at the different Ni sites.
Note that we employ the full tight-binding Hamiltonian based on the \mbox{$\sqrt{2}\times\sqrt{2}$} supercell here instead of a simplified model, describing \textit{explicitly} all Ni sites to accurately treat effects of octahedral rotations and electrostatic doping.
Subsequently, we consider a Kanamori-type interaction vertex during the calculation of $\chi^\text{RPA}_{abcd}(\mathbf{q})$ in the random phase approximation (RPA),
\begin{equation}
    \chi^\text{RPA}(\mathbf{q}) = 
    \left( 1 - \chi^{(0)}(\mathbf{q}) \ \Gamma \right)^{-1}
    \chi^{(0)}(\mathbf{q})   
\end{equation}
with
\begin{equation}
    \Gamma_{abcd} =
    \begin{cases}
    U & \mathrm{if}\;a=b=c=d \\
    U' & \mathrm{if}\;a=c\neq b=d \\
    J & \mathrm{if}\;a=b\neq c=d \\
    J' & \mathrm{if}\;a=d\neq b=c \\
    0 & \mathrm{else}
    \end{cases}
\end{equation}
where $U' = U - 2 J$ and $J' = J$.
Notably, $\Gamma_{abcd}$ characterizes electronic correlations involving orbitals at the same Ni site, vanishing if any indices correspond to distinct sites.
Finally, we implemented a strategy to unfold the result from the \mbox{$\sqrt{2}\times\sqrt{2}$} to the full Brillouin zone by including structural phase factors
and computed the 'physical' susceptibility by evaluating the trace:
\begin{equation}
\chi^\text{RPA}_{}(\mathbf{q}) = \frac{1}{2} \sum_{a,b} \chi^\text{RPA}_{aabb}(\mathbf{q}) \ .
\end{equation}
Leveraging this methodology, we are able to fully consider the effects of octahedral rotations and electrostatic doping that necessitate the use of large supercells, while displaying $\chi^\text{RPA}_{}(\mathbf{q})$ without backfolding for direct comparison with future experiments.

\section*{Data availability}

The data is available upon reasonable request to the authors.

\begin{acknowledgments}
We thank Christine Ah-Yeung, Martin Bluschke, and Steef Smit for valuable discussions on the angle-resolved photoemission spectroscopy of bilayer nickelates.
This work was supported by the National Science Foundation, Grant No.~NSF-DMR-2118718.
\end{acknowledgments}

\section*{Author Contributions}

BG and PJH conceived of the project. JJH, GRS, RGH, and PJH supervised the research. BG performed the theoretical simulations and corresponding analysis. BG and PJH wrote the paper. All authors discussed the results and revised the paper.

\section*{Competing Interests}

The authors declare no competing interests.

\end{document}